\documentclass[letter,longauth]{aa}
\usepackage{natbib}
\usepackage{hyperref}
\hypersetup{backref, colorlinks=true, citecolor=blue}
\usepackage{graphicx}
\usepackage{txfonts}
\begin{document}
   \title{100~$\mu$m and 160~$\mu$m emission as resolved star--formation rate estimators in M33 (HERM33ES)\thanks{{\em Herschel} is an ESA space observatory with science instruments provided by European--led Principal Investigator consortia and with important participation from NASA.}}
   \author{M. Boquien\inst{1} \and D. Calzetti\inst{1} \and C. Kramer\inst{2} \and E. M. Xilouris\inst{3} \and F. Bertoldi\inst{4} \and J. Braine\inst{5} \and C. Buchbender\inst{2} \and F. Combes\inst{6} \and F. Israel\inst{7} \and B. Koribalski\inst{8} \and S. Lord\inst{9} \and G. Quintana-Lacaci\inst{2} \and M. Rela\~{n}o\inst{10} \and M. R\"ollig\inst{11} \and G. Stacey\inst{12} \and F. S. Tabatabaei\inst{13} \and R. P. J. Tilanus\inst{14} \and F. van der Tak\inst{15} \and P. van der Werf\inst{7} \and S. Verley\inst{16}}

   \institute{University of Massachusetts, Department of Astronomy, LGRT-B 619E, Amherst, MA 01003, USA \email{boquien@astro.umass.edu}
   \and Instituto Radioastronomia Milimetrica, Av. Divina Pastora 7, Nucleo Central, E-18012 Granada, Spain
   \and Institute of Astronomy and Astrophysics, National Observatory of Athens, P. Penteli, 15236 Athens, Greece
   \and Argelander Institut f\"ur Astronomie. Auf dem H\"ugel 71, D-53121 Bonn, Germany
   \and Laboratoire d'Astrophysique de Bordeaux, Universit\'{e} Bordeaux 1, Observatoire de Bordeaux, OASU, UMR 5804, CNRS/INSU, B.P. 89, Floirac F-33270, France
   \and Observatoire de Paris, LERMA, 61 Av. de l'Observatoire, 75014 Paris, France
   \and Leiden Observatory, Leiden University, PO Box 9513, NL 2300 RA Leiden, The Netherlands
   \and ATNF, CSIRO, PO Box 76, Epping, NSW 1710, Australia
   \and IPAC, MS 100-22 California Institute of Technology, Pasadena, CA 91125, USA
   \and Institute of Astronomy, University of Cambridge, Madingley Road, Cambridge CB3 0HA, England
   \and KOSMA, I. Physikalisches Institut, Universit\"at zu K\"oln, Z\"ulpicher Stra\ss{}e 77, D-50937 K\"oln, Germany 
   \and Department of Astronomy, Cornell University, Ithaca, NY 14853, USA
   \and Max Planck Institut f\"ur Radioastronomie, Auf dem H\"ugel 69, D-53121 Bonn, Germany
   \and JAC, 660 North A'ohoku Place, University Park, Hilo, HI 96720, USA 
   \and SRON Netherlands Institute for Space Research, Landleven 12, 9747 AD Groningen, The Netherlands
   \and Dept. F\'{i}sica Te\'{o}rica y del Cosmos, Universidad de Granada, Spain 
}

   \date{}

\abstract{Over the past few years several studies have provided estimates of the SFR (star--formation rate) or the total infrared luminosity from just one infrared band. However these relations are generally derived for entire galaxies, which are known to contain a large scale diffuse emission that is not necessarily related to the latest star--formation episode.}{We provide new relations to estimate the SFR from resolved star--forming regions at 100~$\mu$m and 160~$\mu$m.}{We select individual star--forming regions in the nearby (840~kpc) galaxy M33. We estimate the SFR combining the emission in H$\alpha$ and at 24~$\mu$m to calibrate the emission at 100~$\mu$m and 160~$\mu$m as SFR estimators, as mapped with PACS/{\em Herschel}. The data are obtained in the framework of the HERM33ES open time key project.}{There is less emission in the HII regions at 160~$\mu$m than at 100~$\mu$m. Over a dynamic range of almost 2~dex in $\Sigma_{SFR}$ we find that the 100~$\mu$m emission is a nearly linear estimator of the SFR, whereas that at 160~$\mu$m is slightly superlinear.}{The behaviour of individual star--forming regions is surprisingly similar to that of entire galaxies. At high $\Sigma_{SFR}$, star formation drives the dust temperature, whereas uncertainties and variations in radiation--transfer and dust--heated processes dominate at low $\Sigma_{SFR}$. Detailed modelling of both galaxies and individual star forming regions will be needed to interpret similarities and differences between the two and assess the fraction of diffuse emission in galaxies.}


\maketitle

\section{Introduction}

Star formation is one of the main drivers of galaxy formation and evolution and as such the accuracy of the SFR (star--formation rate) determination is of great importance for deriving the cosmic history of galaxies. Along with the UV and the H$\alpha$, the total infrared luminosity is widely used to estimate the SFR. To properly quantify the infrared luminosity a good sampling of the infrared SED (spectral energy distribution) is needed \citep{dale2002a,draine2007a}. Over the past few years, many authors have shown that the TIR (total infrared) luminosity, and by extension the SFR \citep{kennicutt1998a}, can also be evaluated from monochromatic emission measures \citep[][Y. Li et al. 2010, in prep.]{takeuchi2005a,calzetti2007a,rieke2009a,calzetti2010a,boquien2010a}. However, most relations between the far--infrared luminosity and the SFR are established for entire galaxies. At shorter wavelengths, some relations have been derived from individual star--forming regions \citep{calzetti2005a,perez2006a,calzetti2007a,relano2007a}. As new deep surveys will become available at wavelengths where most of the energy is emitted, it is prudent to try to understand the relation between the integral emission of a galaxy and that of the individual star forming regions. The physical conditions, such as temperature, abundance and emissivity, of the infrared--emitting dust can vary widely in a galaxy, and so does the significant contribution from evolved stars \citep{lonsdale1987a,sauvage1992a,buat1996a}. Therefore, any scaling relation established from the emission of entire galaxies may not be appropriate when applied to resolved star--forming regions in these same galaxies.

The limited resolution of far--infrared instruments onboard IRAS (Infrared Astronomical Observatory), ISO (Infrared Space Observatory) or even {\em Spitzer} beyond 60~$\mu$m made the study of individual star--forming regions within galaxies difficult. However, the recently launched {\em Herschel} Space Observatory \citep{pilbratt2010a} with its unprecedented resolution provides the first opportunity to study the spatially resolved far--infrared dust emission in exquisite detail. Such a fine resolution is of the utmost importance to study the emission of star--forming regions located in nearby galaxies in order to provide insights into the fundamental properties of the dust and to quantify the SFR.

With an inclination of 56$^{\circ}$ \citep{regan1994a} and a distance of only 840~kpc \citep{freedman1991a}, M33 is one of the closest spiral galaxies. It has been imaged by {\em Herschel} in the context of the HERM33ES Open Time Key Project \citep{kramer2010a}, providing one of the finest views {\em Herschel} will ever provide of a spiral galaxy from 100~$\mu$m to 500~$\mu$m \citep{kramer2010a,braine2010a,verley2010b}.

\section{Observations and data reduction}
\label{sec:obs}

\subsection{PACS}

\begin{figure*}[!ht]
\includegraphics[width=\columnwidth]{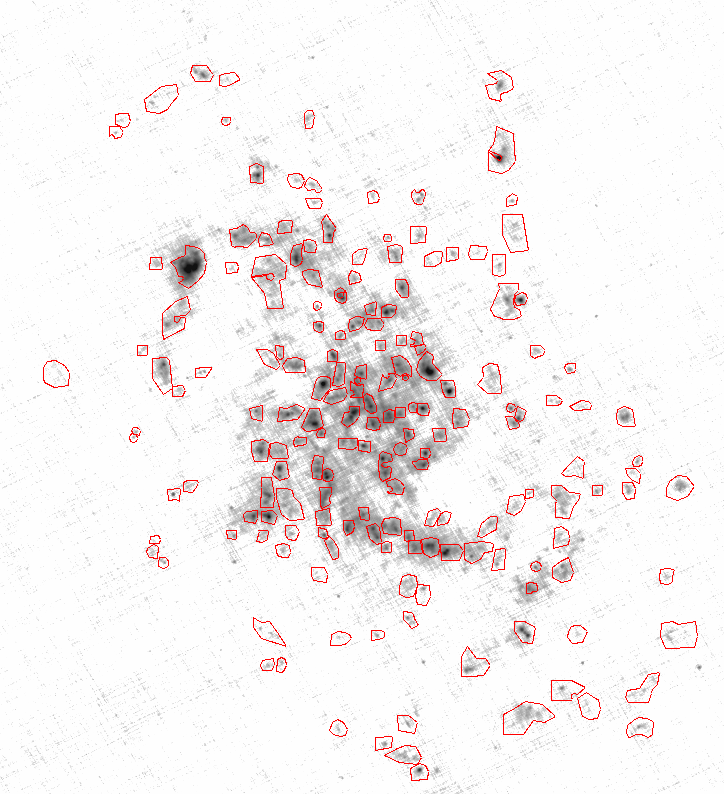}
\includegraphics[width=\columnwidth]{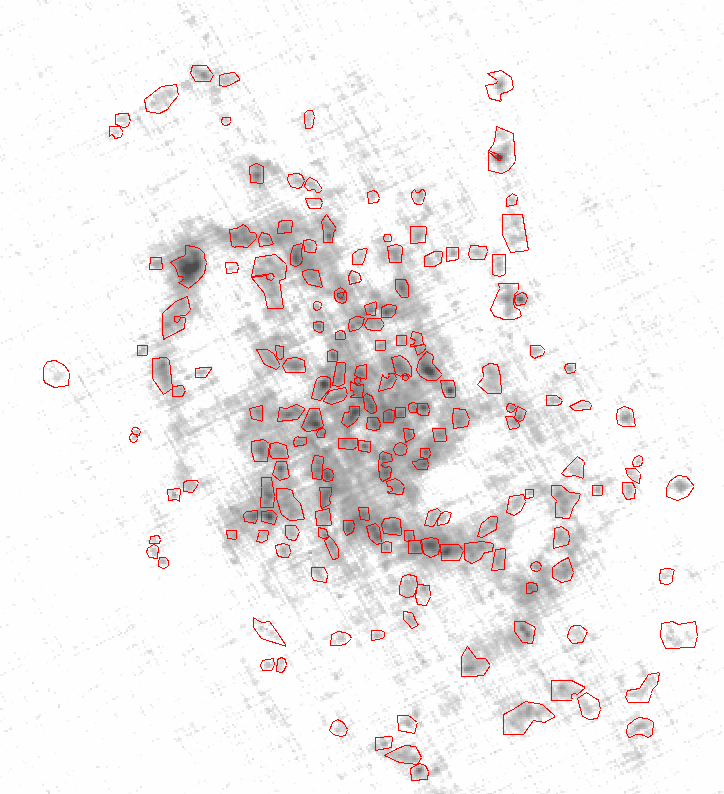}
\caption{PACS maps of M33 at 100~$\mu$m (left) and 160~$\mu$m (right). To probe the full dynamic range of the image a logarithmic scale is used. Individual clumps are well resolved with minimal blending. Some diffuse emission is visible, particularly in the arms and the central region of the galaxy. The visible perpendicular stripes in each scanning direction are sampling artefacts. The red polygons represent the selected apertures. North is up, East is left\label{fig:maps}}
\end{figure*}

The observations provided by PACS \citep{poglitsch2010a} at 100~$\mu$m and 160~$\mu$m are presented by \cite{kramer2010a} along with a detailed description of the data processing pipeline. The observations were carried out on 2010-01-07 in parallel mode with a 20\arcsec/s scanning speed for a total of 6.3 hours, through a single scan and a perpendicular cross scan. The frames were first processed to level 1 with HIPE \citep{ott2010a}, the drifts were corrected and were deglitched with the second--order deglitcher with a 6-$\sigma$ threshold. The maps were produced with photproject mapmaker\footnote{PACS photometer -- Prime and Parallel scan mode release note. V.1.2, 23 February 2010.} using a two--step masking technique to preserve the diffuse emission from being affected by the high--pass filter. The total flux of the galaxy is consistent with the measures provided by IRAS and {\em Spitzer} at 100~$\mu$m and 160~$\mu$m \citep{kramer2010a}. We present the two maps in Figure~\ref{fig:maps}. The  pixel size is 3.2\arcsec\ at 100~$\mu$m and 6.4\arcsec\ at 160~$\mu$m for a spatial resolution of 6.7\arcsec$\times$6.9\arcsec\ at 100~$\mu$m and 10.7\arcsec$\times$12.1\arcsec at 160~$\mu$m. The absolute calibration uncertainty is 5\% at 100~$\mu$m and 10\% at 160~$\mu$m. The total fluxes of M33 agree to within a few percent with those from ISO \citep{hippelein2003a} and MIPS. In addition, for all radial averages the PACS 160 flux is within 20\% of the MIPS 160 one.

\subsection{H$\alpha$ and {\em Spitzer} MIPS 24~$\mu$m}

We used the H$\alpha$ image presented by \cite{hoopes2000a} that is commonly used in the literature \citep{tabatabaei2007a,gardan2007a,verley2007a,verley2009a,verley2010a}. The NII contamination was corrected assuming [NII]/H$\alpha$=0.05 in the filter bandpass. We also corrected the fluxes for Galactic foreground extinction using the \cite{cardelli1989a} law, assuming E(B-V)=0.042 from the NASA Extragalactic Database\footnote{The NASA/IPAC Extragalactic Database (NED) is operated by the Jet Propulsion Laboratory, California Institute of Technology, under contract with the National Aeronautics and Space Administration.}.

We used the 24~$\mu$m MIPS data presented by \cite{tabatabaei2007a}. No further processing was performed on this image.

\subsection{Flux measurements}

All targeted HII regions flux densities were measured in polygonal apertures using IRAF's polyphot procedure. Each polygon was constructed manually from the PACS 160~$\mu$m image and tailored to avoid subtraction artefacts in H$\alpha$ and background sources in MIPS 24~$\mu$m images. Each source was selected to be as compact as possible, taking into account the blending at 160~$\mu$m to avoid the mix of several star--forming regions of different ages and properties. The background was calculated measuring the mode of the pixels distribution in an annulus around the aperture. Annulus pixels falling into the aperture of a source were automatically discarded. The inner radius of the annulus ranges from 20\arcsec\ to 70\arcsec\ by steps of 10\arcsec, which were defined to be larger than the equivalent radius of the aperture: $r=\sqrt{S/\pi}$, where $S$ is the area of the aperture. A scale of 50~pc corresponds to an angular size of 11\arcsec. The width of the annulus was set to 12\arcsec. Aperture correction was performed for 24~$\mu$m data\footnote{Following the formula provided in the MIPS instrument handbook.} and for the PACS bands\footnote{Following the correction provided in PACS photometer -- Prime and Parallel scan mode release note. V.1.2, 23 February 2010.}. As the apertures are not circular, we applied the method presented in \cite{boquien2007a} using the equivalent radius of the aperture.

As a proxy for the SFR we applied the scaling presented by \cite{calzetti2007a}: $\textrm{SFR}=\left[L\left({H\alpha}\right)+0.031L\left(24\right)\right]\times5.35\ 10^{-35}$ in M$_{\sun}$~yr$^{-1}$, where $L\left(H\alpha\right)$ is the H$\alpha$ luminosity in W and, $L\left(24\right)$ is defined as $\nu L_{\nu}$ at 24~$\mu$m in W, assuming a \cite{kroupa2001a} IMF (initial mass function) with a constant SFR over 100~Myr.

For an easier use of the SFR estimator we will provide in this article, we subsequently worked in $\Sigma$ (luminosity surface density) because it is distance--independent, in order to facilitate a comparison with other galaxies. To do so, we divided the luminosity by the area of the polygon measured in kpc$^2$.

\section{Results}
\label{sec:results}
\subsection{General characteristics of HII regions}

We selected a total of 179 HII regions from the 160~$\mu$m map. The physical equivalent radius of the extraction apertures ranges from 37~pc to 256~pc, with a median of 99~pc. $\Sigma_{100}$ ranges from $1.0\times10^{33}$~W~kpc$^{-2}$ to $1.0\times10^{35}$~W~kpc$^{-2}$ and $\Sigma_{160}$ from $1.5\times10^{33}$ W~kpc$^{-2}$ to $5.2\times10^{34}$~W~kpc$^{-2}$. The typical 1-$\sigma$ uncertainties are 0.09, 0.06, 0.02, and 0.03 dex in H$\alpha$, 24~$\mu$m, 100~$\mu$m, and 160~$\mu$m respectively. $\Sigma_{SFR}$ (SFR density) ranges from $1.2\times10^{-3}$~M$_{\sun}$~kpc$^{-2}$~yr$^{-1}$ to $1.5\times10^{-1}$~M$_{\sun}$~kpc$^{-2}$~yr$^{-1}$.

The fraction of the total flux enclosed in the 179 apertures compared to the total flux of M33 is 0.40, 0.43, 0.35, and 0.24 in H$\alpha$, 24~$\mu$m, 100~$\mu$m, and 160~$\mu$m assuming galaxy--integrated fluxes $F(H\alpha)=4.03\times10^{-13}$~W~m$^{-2}$, $F(24)=49.4$~Jy \citep{verley2007a}, $F(100)=1288$~Jy, and $F(160)=1944$~Jy \citep{kramer2010a}. That so little of the total 160~$\mu$m flux associated with the selected HII regions suggests that the large--scale diffuse emission seen at this wavelength may not be directly related to the ongoing massive star--formation, which is consistent with the result of \cite{hinz2004a} for M33, but may be heated by non--ionising B and A stars as observed for instance by \cite{israel1996a} in another galaxy with a similar metallicity, NGC~6822.

\subsection{100~$\mu$m and 160~$\mu$m as SFR estimators}

Being closer to the peak IR emission, the 100~$\mu$m promises to yield an accurate SFR indicator for HII regions and star--formation--dominated galaxies. In this respect, the high angular resolution {\em Herschel} data will be essential to establish the range of applicability and any limitation of this indicator.

First of all $\Sigma_{100}$ and $\Sigma_{160}$ are very well correlated with a Spearman correlation coefficient $\rho_{100-160}=0.92$. This is expected because both bands probe the grey body emission of big grains. It appears that $\Sigma_{100}$ and $\Sigma_{160}$ are also well correlated with $\Sigma_{SFR}$, with a Spearman correlation coefficient $\rho_{SFR-100}=0.85$ and $\rho_{SFR-160}=0.77$ .

\begin{figure*}[!ht]
\includegraphics[width=\columnwidth]{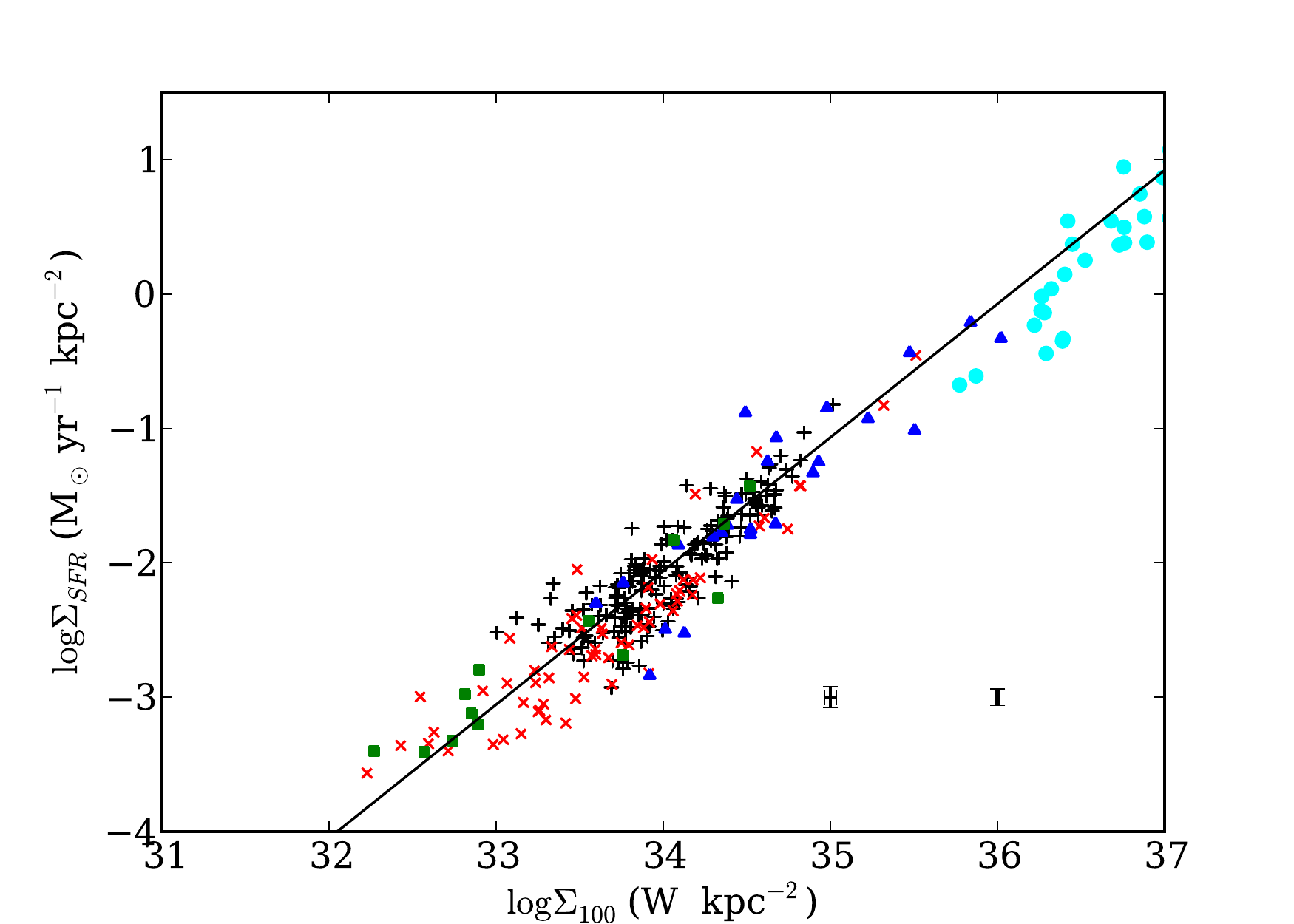}
\includegraphics[width=\columnwidth]{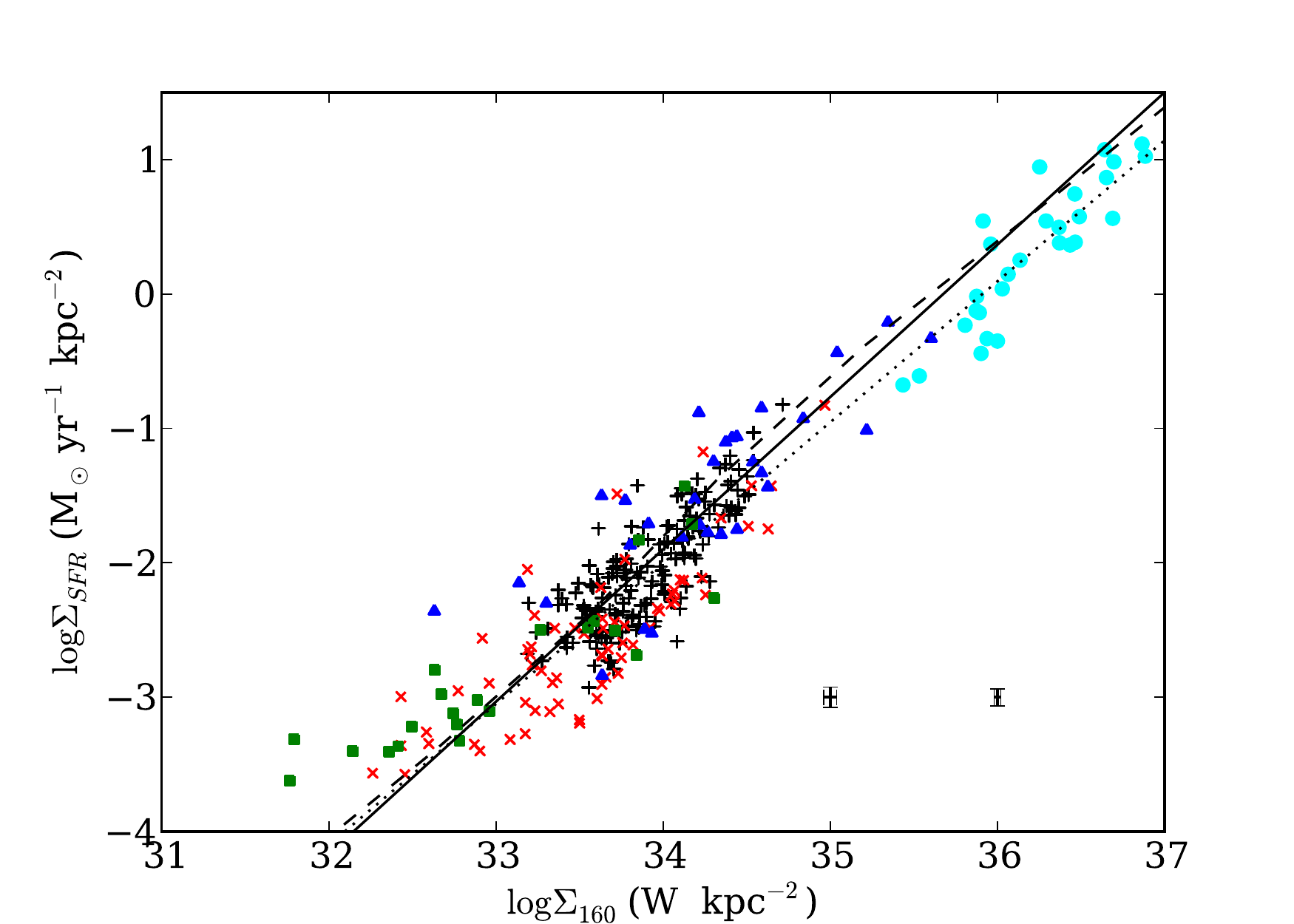}
\caption{$\Sigma_{SFR}$ as a function of $\Sigma_{100}$ (left) and $\Sigma_{160}$ (right). A black plus corresponds to individual HII regions in M33. The error bars correspond to the typical uncertainty for the lower and higher 50\% percentile. For comparison additional integrated galaxies are also plotted: SINGS \citep[{\em Spitzer} Infrared Nearby Galaxies Survey,][red crosses]{dale2007a}, LVL \citep[Local Volume Legacy Survey,][green squares]{dale2009a}, starburst galaxies from \cite{engelbracht2008a} (blue triangles) and LIRGs from \citep[Luminous Infrared Galaxies][cyan circles]{calzetti2010a}. These galaxies belong to the high--metallicity bin of \cite{calzetti2010a}, which encloses the metallicity of M33 \citep{magrini2010a}. The best fit of selected star--forming regions in M33, taking into account the uncertainties on both axes, is plotted with a solid black line. The dashed line represents the model described in Sect. \ref{ssec:modeling} assuming an extinction E(B-V)=0.25, set to reproduce the observed $\Sigma_{H\alpha}/\Sigma_{24}$. Finally the dotted line represents the best fit determined by \cite{calzetti2010a} for entire galaxies.\label{fig:fit-100-160}}
\end{figure*}

In Figure~\ref{fig:fit-100-160} we present the fits of $\Sigma_{100}$ and $\Sigma_{160}$ versus the estimated $\Sigma_{SFR}$. To estimate the relations between the SFR and the PACS emission we fitted a linear relation in log--log using an ordinary least--square technique taking into account uncertainties on both axes.

The best fits for M33 HII regions respectively correspond to
\begin{eqnarray}
\log \Sigma_{SFR}&=&(0.9938\pm0.0244)\log\Sigma_{100}-(35.8520\pm0.8304)\\
\log \Sigma_{SFR}&=&(1.1333\pm0.0293)\log\Sigma_{160}-(40.4325\pm0.9931)
\end{eqnarray}

The relations have a dispersion around the best fit of 0.22 dex and 0.25 dex. We notice that the 100~$\mu$m relation is nearly linear, whereas the 160~$\mu$m relation is slightly superlinear. This is similar to what \cite{calzetti2010a} found (Eq. 23), even though individual star--forming regions probe a much smaller range of $\Sigma_{SFR}$.

\section{Discussion}
\label{sec:discussion}
\subsection{Modeling}
\label{ssec:modeling}

To model the individual HII regions we used the \cite{calzetti2007a} model as a baseline. The ionising flux is determined using {\sc Starburst99} \citep{leitherer1999a} with a \cite{kroupa2001a} IMF, an instantaneous burst and solar metallicity. The extinction is assumed to follow the \cite{calzetti2001a} law. As the $L(H\alpha)/L(24)$ ratio does not show a significant correlation with $\Sigma_{SFR}$ for our selected regions in M33, we assumed a constant E(B-V)=0.25~mag, set to reproduce the observed mean $\Sigma_{H\alpha}/\Sigma_{24}$. For the dust emissivity we assumed the \cite{draine2007a} model prescriptions for {\em Spitzer} MIPS~160, pending updated dust emissivities for {\em Herschel} bands. The model is plotted in Figure~\ref{fig:fit-100-160}.

\subsection{Comparison with entire galaxies}

\cite{calzetti2010a} showed that for entire galaxies the emission at 160~$\mu$m correlates linearly with $\Sigma_{SFR}$. To compare this with the emission of entire galaxies at 100~$\mu$m and 160~$\mu$m, we plotted in Figure~\ref{fig:fit-100-160} several samples of relatively metal--rich galaxies from SINGS ({\em Spitzer} Infrared Nearby Galaxies Survey), LVL (Local Volume Legacy Survey), starburst galaxies from \cite{engelbracht2008a} and some LIRGS (Luminous Infrared Galaxies). Surprisingly, individual star--forming regions and entire galaxies show a similar behaviour. Unfortunately the large scatter in both samples makes any assessment of the contamination by the large diffuse emission in entire galaxies very difficult. However, when they are compared with the relation derived for individual regions, most show a stronger 100~$\mu$m and 160~$\mu$m emission for a given $\Sigma_{SFR}$, especially in the $33\leq\log\Sigma_x\leq34.5$ range where $x$ is 100 or 160. Further comparisons to estimate the contamination by a diffuse far--infrared component will require detailed modelling of individual star--forming regions and entire galaxies.

\subsection{Dust temperature}

\begin{figure}[!ht]
\includegraphics[width=\columnwidth]{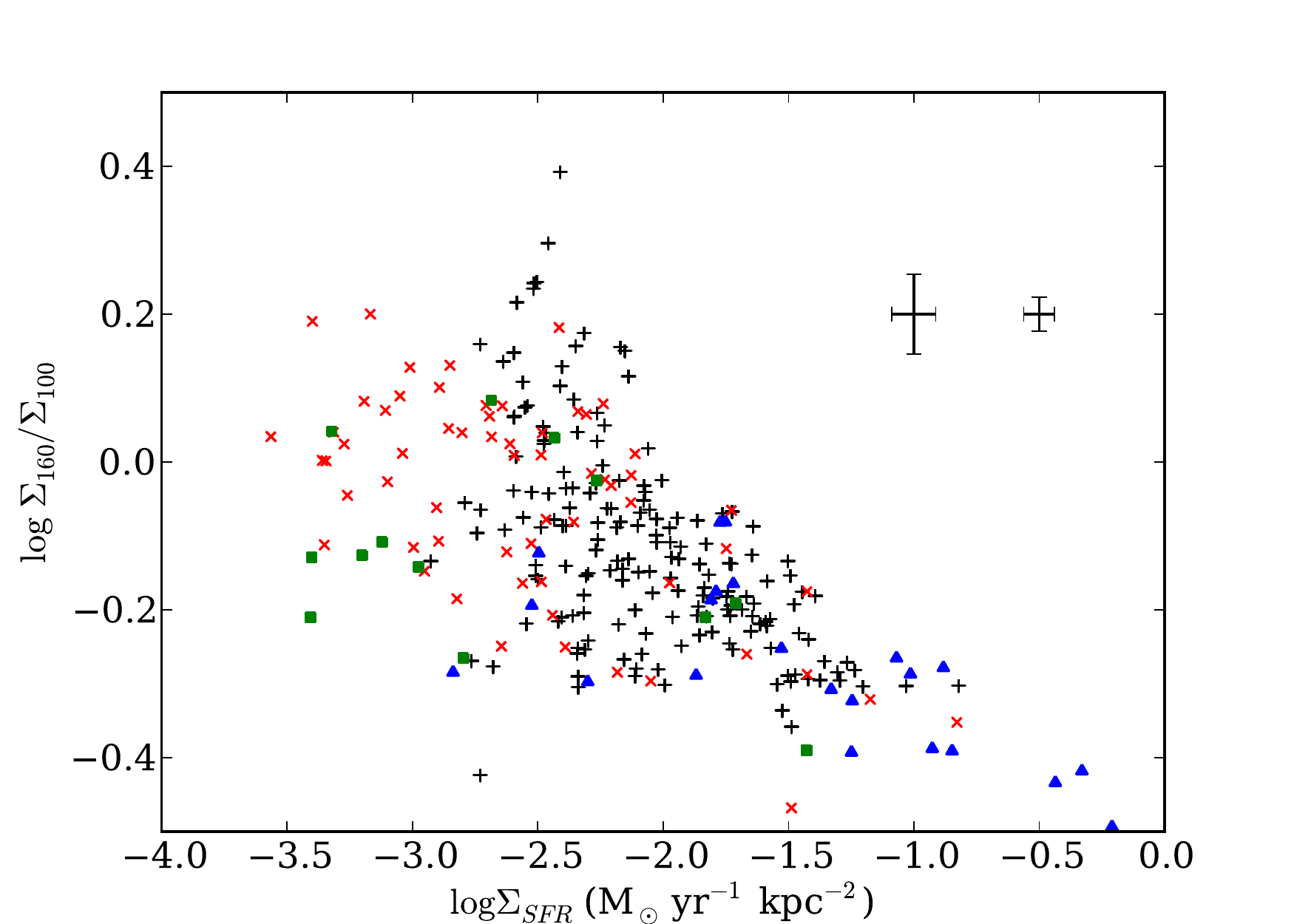}
\caption{$\Sigma_{160}/\Sigma_{100}$ vs. $\Sigma_{SFR}$ (black plusses), for the 179 HII regions selected in M33. The error bars correspond to the typical uncertainty for the lower and higher 50\% percentile. For comparison additional entire galaxies are plotted using the same colour scheme as in Fig.~\ref{fig:fit-100-160}.\label{fig:trend-temp}}
\end{figure}

The slight non--linearity in the $\Sigma_{160}-\Sigma_{SFR}$ relation hints at a higher dust temperature with increasing $\Sigma_{SFR}$, which would have different effects on the two PACS bands because the peak of the emission passes through the filter bandpasses as the temperature increases. Indeed, as dust gets warmer, an increasing fraction will be emitted at shorter wavelengths. In Figure~\ref{fig:trend-temp} we plot $\Sigma_{160}/\Sigma_{100}$, a proxy for the dust temperature of the warm component, versus the $\Sigma_{SFR}$.

We see a clear trend with higher $\Sigma_{SFR}$ ($\rho=-0.55$) leading to a higher dust temperature. The spanned range is compatible with the emissivity values published by \cite{draine2007a} for $0.5\leq U\leq100$, $U$ being the interstellar radiation field normalised to that of the solar neighbourhood. This means that an increasing fraction of the total dust emission is detected in the 100~$\mu$m band compared to the 160~$\mu$m band. Interestingly we also notice that the trend in entire galaxies is very similar to the trend in individual HII regions in M33 despite the fact that individual regions should have little contamination from the diffuse large scale emission. One possible interpretation is that in both entire galaxies and individual HII regions, star--formation dominates at higher $\Sigma_{SFR}$ and creates the trend whereas, at lower $\Sigma_{SFR}$ the trend is influenced by the uncertainties and variations of conditions in the radiation--transfer and dust--heating processes such as the opacity of the star--formation region, the clumpiness of the media, the relative locations of stars and dust clouds, etc.

\section{Conclusions}
\label{sec:conclusion}
We used the high--resolution {\em Herschel} 100~$\mu$m and 160~$\mu$m observations of a nearby star--forming galaxy, M33. We combined {\em Herschel} PACS data with {\em Spitzer} MIPS 24~$\mu$m and ground--based H$\alpha$ to provide new calibrations of the 100~$\mu$m and 160~$\mu$m to estimate $\Sigma_{SFR}$ from individual star--forming regions. For the selected star--forming regions in M33, the 100~$\mu$m luminosity is a linear SFR estimator over a factor 100 in surface brightness, whereas the 160 micron luminosity is slightly superlinear. It appears that individual star forming regions exhibit a similar behaviour as entire galaxies taken from the LVL, SINGS, starburst galaxies from \cite{engelbracht2008a} samples when estimating $\Sigma_{SFR}$ from the 100~$\mu$m and 160~$\mu$m bands emission. In a similar fashion, the dust temperature -- as measured by the ratio of the 160~$\mu$m to 100~$\mu$m emission -- increases as a function of $\Sigma_{SFR}$, suggesting that at high $\Sigma_{SFR}$ the star formation drives the trend for both systems, while at lower $\Sigma_{SFR}$ uncertainties and variations of conditions in the radiation--transfer and dust--heating processes contribute to the scatter. In other words, the fairly wide dust temperature distribution at low $\Sigma_{SFR }$ becomes increasingly biased towards higher temperatures at higher $\Sigma_{SFR}$ in both HII regions and entire galaxies.

\begin{acknowledgements}
We thank {\em Herschel} scientists for their valuable help with the PACS data reduction, in particular Babar Ali, Bruno Altieri, Bidushi Bhattacharya, Nicolas Billot and Marc Sauvage. We also thank the NHSC for providing the computing architecture used in the reduction of the data.
We also thank our referee, C. K. Xu, for useful comments that helped improve the quality of this article.
\end{acknowledgements}
\bibliographystyle{aa}
\bibliography{article}

\end{document}